\documentclass[a4paper,showpacs,superscriptaddress,twocolumn]{revtex4}
\usepackage{epsfig}
\usepackage{graphics}
\usepackage{color}

\def\beq{\begin{equation}}
\def\eeq{\end{equation}}

\begin{document}

\title{Expansion of a Finite Size Plasma in Vacuum }

\author{S. Betti}
\email{betti@df.unipi.it}
\affiliation{Dipartimento di Fisica ``Enrico Fermi'' \& INFM , Universit\`a di
Pisa, Largo B. Pontecorvo 2, 56127 Pisa,
Italy }
\author{F. Ceccherini}
\affiliation{Dipartimento di Fisica ``Enrico Fermi'' \& INFM , Universit\`a di
Pisa, Largo B. Pontecorvo 2, 56127 Pisa,
Italy }
\author{F. Cornolti}
\affiliation{Dipartimento di Fisica ``Enrico Fermi'' \& INFM , Universit\`a di
Pisa, Largo B. Pontecorvo 2, 56127 Pisa,
Italy }
\author{F. Pegoraro}
\affiliation{Dipartimento di Fisica ``Enrico Fermi'' \& INFM , Universit\`a di
Pisa, Largo B. Pontecorvo 2, 56127 Pisa,
Italy }

\date{\today}

\begin{abstract}
The expansion dynamics of a finite size plasma is examined from
    an analytical perspective.
     Results regarding the charge distribution as well as the
electrostatic potential are presented. The acceleration
of the ions and the associated  cooling  of the electrons that
    takes place during the plasma expansion is described.
An extensive analysis of the transition between the semi
infinite  and the finite size plasma behaviour is carried out.
Finally, a test of the analytical results, performed through numerical
simulations, is presented.
\end{abstract}

\maketitle
The production of energetic particles through the interaction of
    an intense laser
pulse with a solid target is a topic that has been widely investigated
during the last
three decades. For sufficiently long pulses, the emitted particles emerge
from a coronal plasma formed by the laser on the  target foil and the expansion
of the plasma in vacuum plays a key role in such a  phenomenon. Thus, despite in
    Refs.\cite{mora, allen1, allen2, lontano} the expansion  of a semi infinite
    plasma has been widely investigated, present
experiments  often involve ``thin foils'' as targets
\cite{borghesi, hegelich, badziak, mackinnon, maksimchuk, clark, snavely} which are
in some
cases no
thicker than a few tens of a Debye length, and we believe that, in
these cases, the experimental results must be analyzed
in terms of the expansion of a finite size plasma.
The finite-size and the semi-infinite cases are very different from each
    other, as in the latter case
an infinite amount of energy is available. As a consequence,
even if the ions are
accelerated,  the energy of the electron plasma
remains constant. On the contrary,
    in the case of a
finite size plasma an exchange of kinetic energy between electrons
and ions takes
place as long as the two populations have different velocity distributions,
    i.e.,
as long as a charge separation is present.
The aim of this work is to provide a detailed analytical
    description of the thermal expansion of a  globally neutral,
finite size, unidimensional plasma. Analytical predictions
   will be compared with numerical results obtained
with a Particle in Cell (PIC) code.\\
The configuration
at the initial time $t_0$, which formally can be defined as
the hydrostatic equilibrium in the limit of infinitely
massive ions, is specified by the ion density
$
n_{i0}(x) \equiv n_i(x, t_0) = n_0~ \theta(a - |x|),
$
   and by the Boltzmann-like electron profile
$
n_{e0}(x) \equiv n_e(x, t_0) = {\bar n}
~ \exp{ \left( e \Phi(x, t_0) / T_{e0} \right) } $, where
$a$ is the half-thickness of the plasma, $2a n_0$ is the total number of
positively charged particle, $\theta(x) = 0$ for $x<0$ and $\theta(x) = 1$
for $x>0$, $\Phi(x, t_0)$ is the
electrostatic potential, ${\bar n}$ is
    the density  at the position
where $\Phi(x, t_0) = 0$, $T_{e0}$ the initial
electron plasma temperature and
    $\int_{-\infty}^\infty n_{e0}(x)\mbox{d}x = 2a n_0$ because the system is
globally neutral.  For the sake of notational simplicity we are considering
    a plasma where the ion
charge is equal and opposite of that of the electrons.
Measuring space in
units of the initial electron Debye length  $\lambda_{d,0}
= (T_{e0}/4 \pi e^2 n_0)^{1/2}$
and rescaling the potential $\Phi(x,t)$ as
$
\Phi(x,t) \to \phi(x,t) =e \Phi(x,t)/T_{e0},
$
we rewrite the Boltzmann equation in the simpler form
\begin{equation}\label{scaled}
n_{e0}(\zeta) = {\bar n} e^{\phi(\zeta)}.
\end{equation}
In order to obtain the electrostatic potential $\Phi(x, t_0)$,
and hence $n_{e0}(x)$,
we need to solve the Poisson equation
$
\nabla^2 \Phi(x, t_0) = -4  \pi e [n_{i0}(x)- n_{e0}(x)],
$
coupled to Eq.(\ref{scaled}).
Because of the  inversion symmetry ($\zeta \to -
\zeta$) of the configuration,  we will restrict our analysis to the half-plane
$\zeta \geq
0$.
We choose the potential to be zero in $\zeta= a$, where $a$ denotes now the
normalized half
width of the plasma slab, so that  ${\bar n} = n_{e0}(a)\equiv n_a$ and
    Poisson
equation reads
\begin{equation}
\label{poisson}
\frac{d^2 \phi}{d\zeta^2}  =  \left( \frac{n_a}{n_0} e^{\phi (\zeta)} - 1 \right)
~
\theta(a - \zeta) + \left( \frac{n_a}{n_0} e^{\phi (\zeta)} \right) ~
\theta(\zeta - a).
\end{equation}
The different  functional form of
$n_{i0}(x)$ for $|x| \leq a$ (internal region) and for  $|x| >
a$ (external region), implies that separate  treatments are required.
First we will calculate the electron  density at the ion  front $x = a$.
Using this result, we will obtain an approximate  analytical expression for the
electrostatic potential in the internal region, while in the external
region we will use
the analytical solution of Eqs.(\ref{scaled}, \ref{poisson}) found in
    \cite{allen1}.\\
The value of $n_a$ can be derived in the following way. Integrating
Eq.(\ref{poisson}) once gives the electric  field $E (\zeta)$ in the form
\begin{equation}
\label{E_int}
E(\zeta)
=  \sqrt{2[(n_a/n_0) \left(e^{\phi(\zeta)} - e^{\phi(0)}
\right) - \left(\phi(\zeta) - \phi(0) \right) ]
}
\end{equation}
for $0 \leq \zeta \le a$ and, for $ \zeta > a$, in the form
\begin{equation}
\label{E_ext}
E(\zeta) =  \sqrt{ 2(n_a/n_0)  e^{\phi(\zeta)} } ,
\end{equation}
where  the integration constants have been fixed such that
in Eq.(\ref{E_int}) $E(0) = 0$, as follows from the inversion  symmetry,
and that  in Eq.(\ref{E_ext}) $
\lim_{\zeta \to \infty}   E(\zeta) = 0$
which expresses charge neutrality
   (together with  $\lim_{\zeta \to \infty} (n_a/n_0)
e^{\phi(\zeta)} = 0$ since there are no electrons at infinity).
Imposing the continuity of $E(\zeta)$ across the boundary $\zeta =  a$
and using Eq.(\ref{scaled}),  we obtain the following
relationship for  the potential and  the
electron density  at $\zeta=0$
$
\phi (0) = (n_a/n_0) e^{ \phi(0) } = n_{e0}(0)/ n_0
$. The integration in
Eq.(\ref{E_int}) for $\phi(x)$ can not be performed explicitly. Numerical
integration
gives the electron density $n_{e0}(0)$ at  $\zeta = 0$ and  the electron density
$n_{a}$ at  $\zeta = a$  as a function of the plasma length $a$. 
%%%%%%%%%%%%%%%
\begin{figure}
\resizebox{0.4\textwidth}{!}{%
 \includegraphics{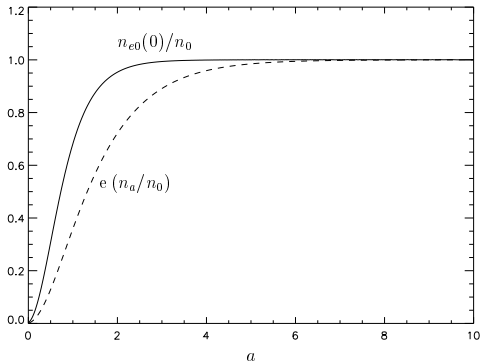}
}
\caption{ Behaviour of $n_a / n_0$ and of $n_{e0}(0) /
n_0$ as a function of $a$.}
\label{density.eps}
\end{figure}
%%%%%%%%%%%%%%% 
Fig. \ref{density.eps} shows that, for $a \gg  1$,  (i.e., in dimensional units
for
$a\gg \lambda_{d,0}$), we find
$\phi (0) = n_{e0}(0)/ n_0 \simeq 1$ and $ n_a/ n_0 \simeq 1/{\rm{e}}$, so that
for most
cases of  interest, the electron density at the ion front is
equal to that obtained in the limit   of a  semi-infinite plasma.
The solution of Eq.(\ref{E_ext}), with the condition $\phi(a) = 0$,
is given by
\begin{equation}
\label{field scaling}
E = \frac{2 \sqrt{n_a/ n_0} }{\sqrt{n_a/ n_0} \left( \zeta - a \right) + \sqrt{2}},
\end{equation}
with electrostatic potential given by
\begin{equation}\label{tt}
\phi (\zeta) = \mathrm{ln} \frac{2}{ \left( \sqrt{n_a/ n_0}~(\zeta - a) + \sqrt{2} \right)^2 }. 
\end{equation}
In order to derive an approximate solution of Eq.(\ref{E_int}),
we define a layer of thickness $\delta$  inside the plasma,
specified by the assumptions  that
${n_{e0}(\zeta)} \simeq {n_{0}}$ for $ 0 < \zeta  < (a-\delta)$,  where
     $E \simeq 0$ and
$\phi \simeq \mathrm{const} = \phi(0)$,
and that  the electron density distribution differs significantly from the
fixed ion distribution only for
     $\zeta > (a - \delta)$. Inside this layer
we adopt an approximate
parabolic fit for the
electrostatic potential of the form
$
\phi(\zeta) =  \mathcal{C}_1 (\zeta - a)  + \mathcal{C}_2 (\zeta - a)^2/2
$,
where   the constant term has been set equal to zero since, by continuity
with the external solution,
$
\phi(a) =0$. This fit of the potential
corresponds to a rough approximation in Poisson's equation (\ref{poisson})
where the electron density inside the
layer is taken to be constant and equal to an intermediate value between its
two values, $n_a$ and $n_0$, at the borders
of the layer. This value must be determined,
together with
$C_1$ and $\delta$,  self-consistently by requiring
that  the potential  and the electric field be
continuous  at $\zeta =a, a-\delta$, which
ensures in particular charge conservation.
    From
the continuity of the electric field at $\zeta = a$ we obtain
from Eq.(\ref{field scaling})
$\mathcal{C}_1 = - \sqrt{2 / {\rm e} }$.
At $\zeta = a-\delta$, the continuity of the electric field gives
$\mathcal{C}_2 = - \mathcal{C}_1 / \delta$,
while the continuity of $\phi$ defines $\delta$ as  $\delta =
\sqrt{2 {\rm e} } \phi_0  \approx \sqrt{2 {\rm e} }$.\\
We now derive a physical model capable of describing
the electron cooling that occurs as the plasma expansion takes place and,
consequently, ion acceleration.
Measuring temperatures in units of $T_0$, where $T_0 = T_{e0}$,
    mass in units of the ion mass $m_i$, and
introducing the dimensionless time unit $\tau = \omega_{\rm pi} t$,
with $\omega_{\rm pi} = \sqrt{4 \pi n_0 e^2}$,
   we denote with $T_e$ the electron time
dependent temperature,
with $m_e$ the electron mass and
assume that both the initial fluid and thermal ion energy
is zero. The quantity $T_e$ can in principle depend both on $x$ and $\tau$,
   however in
what follows it will be assumed spatially constant.
In order to derive an analytical expression for
the cooling of the electron population we need
to introduce a simplified description of the
ion expansion and of the time evolution of the
electric field. 
%%%%%%%%%%%%%%%
\begin{figure}
\resizebox{0.4\textwidth}{!}{%
 \includegraphics{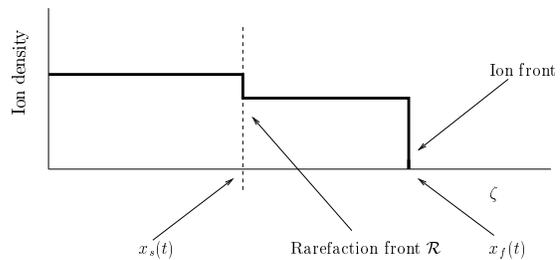}
}
\caption{Model profile of the ion density as plasma
expands.}
\label{ion.eps}
\end{figure}
%%%%%%%%%%%%%%%
For this purpose, we introduce the approximate
ion profile $n_i(x,\tau)$ sketched in Fig.(\ref{ion.eps})
which is specified once the quantities $x_s$ and $x_f$ are known.
The point $x_s$ models the position of the rarefaction
front, which obeys the equation of motion
\begin{equation}
\label{x_s}
x_s = a - \delta - \int_0^{\tau}{c_s d \tau}.
\end{equation}
with $c_s$ the time dependent ion acoustic speed which,
    in dimensionless units, is given by $c_s =  \sqrt{T_e}$.
The quantity $x_f$ represents the ion front position, defined by
    the condition $n_i(x, \tau) = 0$ for $x > x_f$, which obeys
the equation of motion
\begin{equation}
\label{x_f}
\ddot{x}_f = E_{f},
\end{equation}
with initial conditions $\dot{x}_f |_{\tau = 0} = 0$, $x_f |_{\tau = 0} = a$
and $E_{f}$ the self consistent electric field at point $x = x_f$.
Profile (\ref{ion.eps}) then becomes
$
n_i(x, \tau) =  n_0 ~ \theta(x_s - |x|) + n_0 D ~ \theta(|x| - x_s)
- n_0 D ~ \theta(|x| - x_f)
$,
with $D = (a - x_s)/(x_f-x_s)$, which corresponds to
redistributing the charge $n_0 \left( a-x_s \right)$
on the interval $\left( x_f-x_s \right) ~ $\footnote{Note that this
profile can be
    used only if $x_s \geq 0$.}.
Assuming hydrostatic equilibrium at every time $\tau$
between the electron and
ion plasma we find $~ D ~ n_0 / n_f = \rm{e} ~ $ for
the electron density $n_f$ at point $x = x_f$,
thus $E_f = \sqrt{ \left( 2 / {\rm e} \right) D T_e}$,
which completes the description of both the ion profile and dynamics.
Regarding the electrons, we assume quasineutrality in the whole region
$[0, x_f - \delta]$ except for a small interval $[x_s - \Delta x, x_s]$, with
$\Delta x$ to be determined, in
which charge separation occurs such that Boltzmann relation
    (\ref{scaled}) can be satisfied.
The electric field $E_{cs}$
in $[x_s - \Delta x, x_s]$  is approximated by
$E_{cs} \simeq \left( 1 - D \right)$, which is the field of a plane
    capacitor with surface charge density
$\sigma = \left( 1 - D \right) / 4 \pi$ in dimensionless units.
Therefore $-  \int_{x_s - \Delta x}^{x_s} E_{cs} dx = T_e \left[ \phi \left( x_s \right)
     - \phi \left( x_s - \Delta x  \right) \right]$ which defines
$\Delta x$ as $\Delta x = - T_e (\ln{D})/(1-D)$.
The cooling phenomenon is associated with the reflections that the electrons
undergo at  the potential barriers in the region $[x_s - \Delta
x, x_s]$ ({\it internal} barrier) and $[x_f - \delta,+\infty]$
({\it external} barrier), which are moving with velocities
$-c_s$ and $v_f$, respectively. Assuming the reflections
to be elastic, $m_e \ll 1$ implies that the power loss due to a
reflection of an electron with velocity $v$ is given by
$
\Delta U (v) =  2m_e \left( v - \bar{v} \right) \bar{v}
$,
with $\bar{v} = \{ - c_s, ~ v_f \}$.
   The corresponding power is given by $\Delta  U(v)$
   times the number of electrons with velocity $v >
\bar{v}$ that hit the barrier
    per unit time, integrated over the electron distribution function.
    Assuming the latter to be Maxwellian,
the number of electrons with velocity in the interval $[v, v+dv]$
reflected per unit time by the internal and external barrier
is given by $(1-D) n_0 v f(v)$ and $D n_0 v f(v) ~ \theta(v-v_f)$
respectively, where $f(v) = \left( 1 / \sqrt{2 \pi} ~ v_{th} \right)
    e^{- v^2 / 2 v_{th} } $ and $v_{th} = \sqrt{T_{e}/m_e}$.
The total energy of the electron plasma is $U = n_0 a T_e/2$, hence, taking the
temperature $T_e$ to be uniform, $ d U /d \tau
=  (n_0 a/2)  ~{d T_e}/{d \tau}$ and,
taking $v_f, ~ c_s \ll v_{th}$, we find
\begin{equation}\label{cooling law}
\frac{1}{T_e} \frac{ d T_e }{d \tau}
= - \left( 2 / a \right) \left[D v_f-\left(1-D\right) c_s\right],
\end{equation}
which is the electron cooling equation, to be coupled both to Eq.(\ref{x_s})
and Eq.(\ref{x_f}) and then integrated numerically.
The time dependent total kinetic energy $U_i$ of the ions
satisfies  the conservation equation $U_i = n_0 a / 2 - n_0 a T_e/2 - U_{el}$, where $U_{el}$
is the contribution from the release of the electrostatic energy present in the
   initial ``equilibrium'' configuration described previously.
Note that $2 U_{el} / (T_e) \approx  \left(2d U_{el} / dt \right) / \left(
   d T_e/ dt \right)   \approx
\delta / a$ and thus  $U_i \approx 1 - T_e/2$ except for the very beginning
    of the expansion process, when the electrostatic energy
contribution is significant and must be taken
into account.\\
We may  summarize the  results of the above modeling of the plasma expansion in
vacuum by referring  to $a$, $\tau / (a \sqrt{m_e})$ and $\tau / \sqrt{a}$
   as the
relevant physical quantities characterizing the plasma dynamics.
The condition $a \gg 1$ implies  that, at every time $\tau$,
the electron density at the expanding front
is  $n_f = D n_0 / {\rm e}$, as in the  hydrodynamic equilibrium of a
semi-infinite slab. The quantity
$\tau / (a \sqrt{m_e})$, where $a \sqrt{m_e}$ is the
typical time it takes an electron to cross the target,
roughly gives the ratio of particles which, at time
$\tau$, have undergone at least one reflection at the ion front.
Thus the initial expansion phase when $\tau / (a \sqrt{m_e}) \ll 1$
corresponds to the semi-infinite, constant temperature
plasma limit discussed in \cite{mora, allen1, allen2}.
The quantity $\tau / \sqrt{a}$ measures, at the early stage of the
process,  the expansion of the ion component in the plasma. In fact, taking
$v_f \propto \tau$, which corresponds to a uniformly accelerated motion of
the front and neglecting coefficients of order unity,  we obtain $(x_f -a)/a
= dV/V \approx \tau^2/a$ where $V$ is the volume occupied by the  ions. In
the initial expansion phase,  from  Eq.(\ref{cooling law}) we obtain
${d T_e}/{T_e} \approx - 2 ~ (v_f \tau ) / a \approx - 2 ~{dV}/{V}$,
which implies that in this phase the plasma behaves like a perfect
   monoatomic gas obeying the law of unidimensional
adiabatic expansion.
%%%%%%%%%%%%%%%
\begin{figure}
\resizebox{0.4\textwidth}{!}{%
 \includegraphics{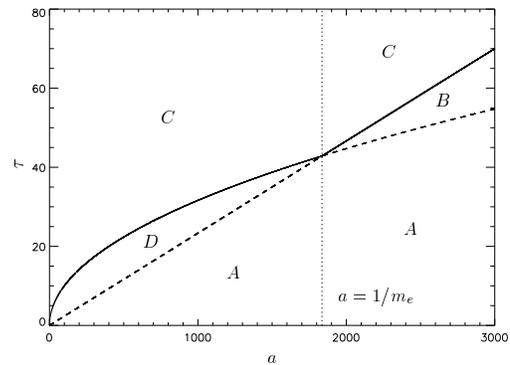}
}
\caption{\label{parameter.eps} Plasma expansion regimes in 
the $(a,\tau)$-plane.
Electron cooling is important in region C. The  uniform electron 
temperature assumption  used
in our analytical model  applies for $a< 1/m_e$.}
\end{figure}
%%%%%%%%%%%%%%%
 %%%%%%%%%%%%%%
\begin{figure}
\resizebox{0.4\textwidth}{!}{%
 \includegraphics{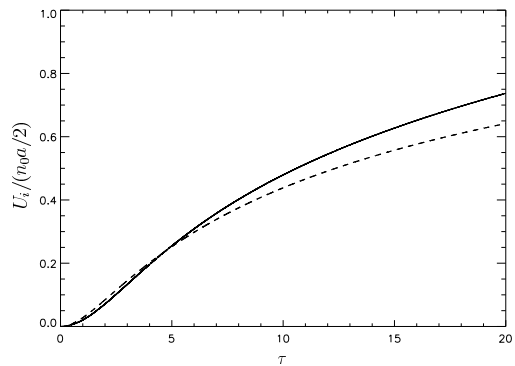}
}
\caption{\label{T.eps} Total normalized ion kinetic energy versus time $\tau$
as obtained from the analytical model (dashed line) and from  a PIC simulation
(solid line) in a plasma with $a = 50 ~ \lambda_d$.}
\end{figure}
%%%%%%%%%%%%%%%
%%%%%%%%%%%%%%
\begin{figure}
\resizebox{0.4\textwidth}{!}{%
 \includegraphics{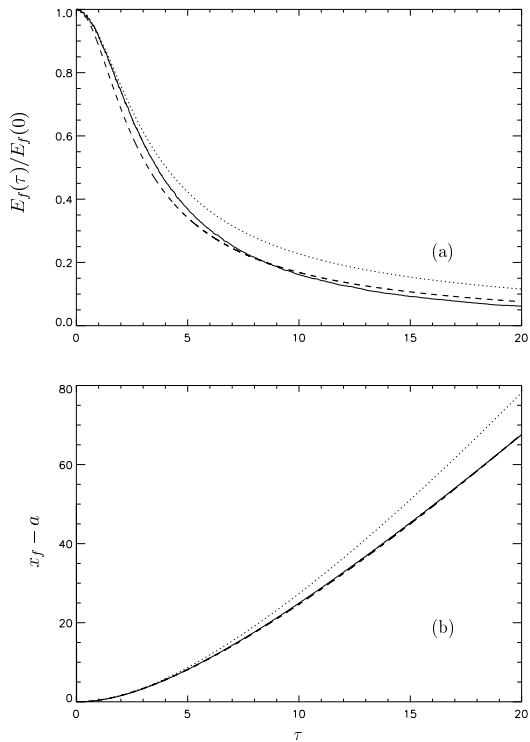}
}
\caption{\label{confronto.eps}  Normalized electric field $E_f$
at the ion front (a)  and  ion front position $x_f$  (b) versus 
time $\tau$  as obtained
from the analytical
  model (dashed line) and
from  the PIC simulation (solid line),
in a plasma with $a = 50 ~ \lambda_d$.
The dotted line corresponds to semi-infinite case \cite{mora}.}
\end{figure}
%%%%%%%%%%%%%%%
Thus, the plane $(a, \tau)$ can be divided  into four regions by the
   two curves $\tau = \sqrt{a}$ and $\tau = a \sqrt{m_e}$,
   as shown in Fig.(\ref{parameter.eps}). The two curves cross at $a =1/m_e$.
In the three regions $A, B, D$ defined by
$\tau < max (a \sqrt{m_e} ,\sqrt{a})
$, electron cooling is not important and the slab expansion can be
approximated by the semi infinite  case.
In region  $C$ the effects of the
the finite size of the plasma becomes apparent and  ion acceleration
takes place together with electron cooling.\\
Now we reexamine the approximations
used  in the derivation of Eq.(\ref{cooling law}).
The assumption  of uniform
temperature rests on the comparison between
the energy redistribution time, approximated by $a \sqrt{m_e}$,
and the typical ion plasma expansion time
$\sqrt{a}$, which roughly represents the energy
  confinement time. Therefore the temperature
inside the plasma can be taken to  be uniform
as long as $a m_e \ll 1$ (the area to the left of the vertical dotted line in
Fig.(\ref{parameter.eps})) and  the predictions of our analytical model
are thus restricted to this domain which, however,
covers the typical experimental configurations
\cite{mackinnon, snavely}.  The  approximation that the electron
reflections are elastic in the frame   co-moving with the reflecting barrier,
can be written  as
$\sqrt{m_e} {\partial}[  \ln{(
T_e \phi )}] /{\partial \tau} \ll 1$, where we
take  the electron reflection time  to be of order $\sqrt{m_e}$.
The validity of this condition has been verified  by integrating
Eqs.(\ref{x_s}, \ref{x_f}, \ref{cooling law}) numerically. Finally,
because of the significant spread of the ion front, the effective
value of $E_f$ is smaller than the one obtained within this model.
This may be corrected by introducing  the   reduction factor
   $ 1 - \left( 1 - D \right)^\alpha$, with $\alpha$ a constant to be fixed
numerically.
In order to determinate the value of  $\alpha$ and to verify
our  analytical results,  a series of numerical simulations has
been performed using a  unidimensional PIC code.
The numerical results that we present here describe the expansion of
  a plasma slab  $50 ~ \lambda_d$ thick with
initial conditions given by Eqs.(\ref{scaled},\ref{tt}).
  An overall  good agreement between the analytical and
  the numerical results is achieved with  $\alpha =
1.4$. The ion  total kinetic
  energy  and the
value of $E_f$ and the position of the ion front  are plotted versus time in
Figs.(\ref{T.eps}) and (\ref{confronto.eps}),
respectively.  Different simulations for slabs up to $150 \lambda_d$
thick have been performed and have
  confirmed the validity of the
  analytical model and in particular of the choice of the value
  of the parameter  $\alpha$.   The comparison  between the finite size
results  and those described  in \cite{mora} is shown in
Fig.(\ref{confronto.eps}) and indicates that the semi-infinite and
  the finite size plasma behaviours are
very similar only at the early stage of the expansion process,
while they become, as expected,  substantially  different when
  the electron  kinetic energy is
significantly reduced.\\
In conclusion  our analytical model  shows that, in the case  of a finite size
plasma,   the  process of electron cooling leads to a nearly complete
energy transfer  from electrons to ions  on  time intervals of the order of a few
tens of the  transit time $a \sqrt{me}$. This sets
  an upper limit to
the  energy ions can acquire in the thermal  expansion of a finite size plasma.
Furthermore, different expansion regimes as function of time and plasma size have
been identified  and the  numerical verification of the proposed model
has been discussed.\\
 This work  was supported by the INFM Parallel Computing Initiative.

\end{document}